\documentclass{article}
\usepackage{graphicx}
\usepackage{here}
\def\beq{\begin{equation}}
\def\eeq{\end{equation}}
\def\bea{\begin{eqnarray}}
\def\eea{\end{eqnarray}}
\def\bq{\begin{quote}}
\def\eq{\end{quote}}

\def\bl{\bullet}
\def\nnb{\nonumber}
\def\ga{\left(}
\def\dr{\right)}
\def\aga{\left\{}
\def\adr{\right\}}
\def\lb{\lbrack}
\def\rb{\rbrack}

\def\rar{\rightarrow}
\def\nnb{\nonumber}
\def\la{\langle}
\def\ra{\rangle}
\def\nin{\noindent}
\def\ba{\begin{array}}
\def\ea{\end{array}}
\def\bm{\overline{m}}
\begin{document}
\topmargin -2.cm
\oddsidemargin -.5cm
\evensidemargin -1.0cm
\pagestyle{empty}
\begin{flushright}
\end{flushright}
\vspace*{1cm}
\begin{center}
\section*{Strange quark mass 
from    $e^+e^-$ revisited \\ and present status of light quark masses }
\vspace*{0.5cm}
{\bf Stephan Narison} \\
\vspace{0.3cm}
Laboratoire de Physique Th\'eorique et Astrophysiques,
Universit\'e de Montpellier II\\
Place Eug\`ene Bataillon\\
34095 - Montpellier Cedex 05, France\\
\vspace*{1.5cm}
{\bf Abstract} \\ \end{center}
\vspace*{2mm}
\noindent
We reconsider the determinations of the strange quark mass $m_s$ from $e^+e^-$ into hadrons
data using a new combination of FESR and revisiting the existing $\tau$-like sum rules by including
non-resonant contributions to the spectral functions. 
To order $\alpha_s^3$ and including the tachyonic gluon mass $\lambda^2$ contribution, 
which phenomenologically parametrizes the UV renormalon
effect into the PT series, we obtain the invariant mass 
$\hat m_s=(119\pm 17)$~MeV leading to: $\overline{m}_s$(2~GeV)=($104\pm 15$)~MeV.
Combining this value with the recent and independent phenomenological determinations from
some other channels, to order $\alpha_s^3$ and including $\lambda^2$ , we deduce the
weighted average:  $\overline{m}_s$(2 GeV)=($96.1\pm 4.8$)~MeV .
The positivity of the spectral functions in the (pseudo)scalar [resp. vector]
channels leads to the lower [resp. upper] bounds of $\bm_s(2~\rm{ GeV})$:
$(71\pm 4)~{\rm MeV}\leq\bm_s(2~\rm{ GeV})\leq (151\pm 14)$~MeV, to order $\alpha_s^3$.
Using the ChPT mass ratio
$r_3\equiv 2m_s/(m_u+m_d)=24.2\pm 1.5$, and the average value of $m_s$, we deduce:
 $(\bm_u+\bm_d)$(2 GeV)=($7.9\pm 0.6$)~MeV, consistent with the pion sum rule result, which, combined with the
ChPT value for
$m_u/m_d$, gives:
$\bm_d$(2 GeV)=($5.1\pm 0.4$)~MeV and
$\bm_u$(2~GeV)=($2.8\pm 0.2$)~MeV.  Finally, using
$(\bm_u+\bm_d)$ from the pion sum rule and the average value of $\bm_s$ (without the pion sum rule), the method
gives:
$r_3= 23.5\pm 5.8$ in perfect agreement with the ChPT ratio, indicating the self-consistency of the sum rule results.
Using the value: $\bm_b(\bm_b)=(4.23\pm 0.06)$ GeV, we also obtain the model-building useful scale-independent mass ratio: $m_b/m_s=50\pm 3$.
\vspace*{3.0cm}
\begin{flushleft}
\end{flushleft}
\vfill\eject
\setcounter{page}{1}
 \pagestyle{plain}
\section{Introduction} \par
The determination of the strange quark mass  is of prime
importance for low-energy phenomenology, for CP-violation
and for beyond standard model-buildings.
Since
the advent of QCD, where a precise meaning for
the definition of the running quark masses within 
the $\overline{MS}$-scheme \cite{FLO} has been provided,
a large number of
efforts have been devoted to the determinations of the strange quark mass \footnote{For reviews, see e.g.
\cite{GASSER,RAFAEL,SNL,SNB}.}
using QCD spectral sum rules (QSSR) \footnote{For a review see e.g. \cite{SNB}.} \`a la
SVZ \cite{SVZ}, in the pseudoscalar
\cite{PSEUDO,PSEUDO2,SNL,SNB,MALT2,CHET2}, the scalar
\cite{SCAL,OLLER}, the $e^+e^-$ \cite{GMO,REIND,SNE,MALT,JAMINE,SNL,SNB} channels, 
tau-decay data \cite{PICH,KUHN,SNTAU} and lattice simulations
\cite{LATT,LATT2,LATT3,ISHIKAWA}, while some bounds have been also derived from the positivity of the spectral functions
\cite{BOUND,SNE,SNL,SNB} and from the extraction of the quark condensate \cite{BOUND2,SNL,SNB}.
\\
\nin
In the following, we reconsider the determinations of $m_s$ from
$e^+e^-$ into hadrons data by using new combinations of FESR \cite{SNTAU} and by revisiting the analysis done
in \cite{SNE}. In so doing, we take into account more carefully the small non-resonant contributions into the
spectral functions, though negligible in the individual sum rules become important in the combinations sensitive
to leading order to $m_s$. We also present a new combination of sum rule used in \cite{SNTAU} that we confront with 
previous sum rules presented in \cite{SNE}. We conclude the paper by a comparison of recent 
different determinations of $m_s$ from QCD spectral sum rules from which we extract the average. This average being
confronted to lattice calculations.
\section{Normalizations and notations}
We shall be concerned with the transverse two-point correlator:
\beq
\Pi^{\mu\nu}_{ab}(q)\equiv i\int d^4 x~ e^{iqx}\la 0|{\cal T} J^\mu_a(x)\ga
J^\nu_b(0)\dr^{\dagger}|0\ra=-(g_{\mu\nu}q^2-q_\mu q_\nu)\Pi_{ab}(q^2)~,
\eeq
built from the $SU(3)$ component of the local electromagnetic current:
\beq
J^\mu_{EM}=V^\mu_3(x)+\frac{1}{\sqrt{3}}V^\mu_8(x)~,
\eeq
where:
\beq
V^\mu_a(x)\equiv \sqrt{\frac{1}{2}}\bar{\psi}(x)\lambda_a\gamma^\mu {\psi}(x)~; 
\eeq
$\lambda_a$ are the diagonal flavour $SU(3)$ matrices:
\beq
\lambda_3=\sqrt{\frac{1}{2}}\ga
\ba{ccc}
1&&\\
&-1&\\
&&0\\
\ea
\dr ~,\,\,\,\,\,\,
\lambda_8=\sqrt{\frac{1}{6}}\ga
\ba{ccc}
1&&\\
&1&\\
&&-2\\
\ea
\dr~,
\eeq
acting on the basis defined by the up, down and strange quarks:
\beq
\psi(x)=\ga
\ba{c}
u(x)\\
d(x)\\
s(x)\\
\ea
\dr~.
\eeq
In terms of the diagonal quark correlator:
\beq
\Pi^{\mu\nu}_{jj}(q)\equiv i\int d^4 x e^{iqx}\la 0|{\cal T} J^\mu_j(x)\ga
J^\nu_j(0)\dr^{\dagger}|0\ra=-(g_{\mu\nu}q^2-q_\mu q_\nu)\Pi_{jj}(q^2)~\, \,
j=u,d,s~,
\eeq
where $J^\mu_j(x)\equiv \bar\psi_j\gamma^\mu\psi_j$, the previous $SU(3)$ flavour components of the electromagnetic correlator 
read\footnote{We shall follow the normalization
used in SN \cite{SNE}.}:
\bea
\Pi_{33}&=&\frac{1}{2}.\frac{1}{2}\ga \Pi_{uu}+\Pi_{dd}\dr~,\nnb\\
\Pi_{88}&=&\frac{1}{2}.\frac{1}{6}\ga \Pi_{uu}+\Pi_{dd}+4\Pi_{ss}\dr~,\nnb\\
\Pi_{38}&=&\frac{1}{4\sqrt{3}}\ga \Pi_{uu}-\Pi_{dd}\dr~,\nnb\\
\eea
Therefore, the $e^+e^-\rar$ hadrons total cross-section reads:
\beq
\sigma(e^+e^-\rar{\rm hadrons})_{u,d,s}=\frac{4\pi^2\alpha}{s}e^2\frac{1}{\pi}
\aga \rm{Im} \Pi_{33}(s)+\frac{1}{3}\rm{Im} \Pi_{88}(s)+\frac{2}{\sqrt{3}}\rm{Im}
\Pi_{38}(s)\adr~.
\eeq
In a narrow-width approximation (NWA), the resonance $H$ contributions to the
spectral functions can be introduced through:
\beq
\la 0|V^\mu_a|H\ra=\epsilon^\mu\frac{M_H^2}{2\gamma_{Ha}}~,
\eeq
where the coupling $\gamma_{Ha}$ is related to the meson leptonic width as:
\beq
\Gamma_{H\rar e^+e^-}=\frac{2}{3}\alpha^2\pi\frac{M_H}{2\gamma_{Ha}^2}~,
\eeq
which is itself related to the total cross-section:
\beq
\sigma(e^+e^-\rar H)=12\pi^2\frac{\Gamma_{H\rar e^+e^-}}{M_H}\delta (s-M^2_H)~.
\eeq

\section{QCD corrections and RGI parameters}
In order to account for the radiative corrections,
one introduces the expressions of the running coupling and masses. \\
$\bl$ To three-loop accuracy, the running
coupling can be parametrized as \cite{BNP,SNB}:
\bea
a_s(\nu)&=&a_s^{(0)}\Bigg\{ 1-a_s^{(0)}\frac{\beta_2}{\beta_1}\log
\log{\frac{\nu^2}{\Lambda^2}}\nnb \\
&+&\ga a_s^{(0)}\dr^2\lb\frac{\beta_2^2}{\beta_1^2}\log^2\
\log{\frac{\nu^2}{\Lambda^2}}-\frac{\beta_2^2}{\beta_1^2}\log
\log{\frac{\nu^2}{\Lambda^2}}-\frac{\beta_2^2}{\beta_1^2}
+\frac{\beta_3}{\beta_1}\rb+{\cal{O}}(a_s^3)\Bigg\},
\eea
with:
\beq
a_s^{(0)}\equiv \frac{1}{-\beta_1\log\ga\nu/\Lambda\dr}
\eeq
and
$\beta_i$ are the  ${\cal{O}}(a_s^i)$ coefficients of the
$\beta$-function in the $\overline{MS}$-scheme, which read for three flavours \cite{SNB}:
\beq
\beta_1=-9/2,~~~~\beta_2=-8,~~~~\beta_3=-20.1198.
\eeq
$\bl$ The expression of the running quark mass in terms of the
invariant mass $\hat{m}_i$ is \cite{FLO,SNB}:
\bea\label{eq: mshat}
&&\bm_i(\nu)=\hat{m}_i\ga -\beta_1 a_s(\nu)\dr^{-\gamma_1/\beta_1}
\Bigg\{1+\frac{\beta_2}{\beta_1}\ga \frac{\gamma_1}{\beta_1}-
 \frac{\gamma_2}{\beta_2}\dr a_s(\nu)\nnb \\
&&+\frac{1}{2}\Bigg{[}\frac{\beta_2^2}{\beta_1^2}\ga \frac{\gamma_1}
{\beta_1}-
 \frac{\gamma_2}{\beta_2}\dr^2-
\frac{\beta_2^2}{\beta_1^2}\ga \frac{\gamma_1}{\beta_1}-
 \frac{\gamma_2}{\beta_2}\dr+
\frac{\beta_3}{\beta_1}\ga \frac{\gamma_1}{\beta_1}-
 \frac{\gamma_3}{\beta_3}\dr\Bigg{]} a^2_s(\nu)+
1.95168a_s^3(\nu)\Bigg\},
\eea
where $\gamma_i$ are the ${\cal{O}}(a_s^i)$ coefficients of the
quark-mass anomalous dimension, which read for three flavours \cite{SNB}:
\beq
\gamma_1=2,~~~~\gamma_2=91/12,~~~~\gamma_3=24.8404.
\eeq
$\bl$ The perturbative expression of the correlator reads, in terms of the running coupling evaluated
at $Q^2=\nu^2$ \cite{SNB}:
\beq
-Q^2{d\over dQ^2}\Pi_{ss}(Q^2)= {1\over 4\pi^2}\Bigg{\{}1+\ga a_s\equiv 
\frac{\alpha_s(Q^2)}{\pi}\dr+1.6398
a_s^2-\Bigg{[} 10.2839-\ga{\beta_1^2\over 4}\dr\ga{\pi^2\over 3}\dr\Bigg{]} a_s^3+...\Bigg{\}},
\eeq
where the last extra term in the $a_s^3$ coefficient compared with the expression of the spectral function Im $\Pi$ comes from the analytic
continuation.\\
$\bl$ The $D=2$ contribution reads to order $\alpha_s^3$, in terms of the running mass and by including the  
tachyonic gluon mass $\lambda^2$ term \cite{BNP,CNZ,SNB,CHETKUHN}:
\beq\label{eq:o2}
 Q^2\Pi_{ss}^{(D=2)}(Q^2)\simeq -{1\over 4\pi^2}\Bigg{\{}1.05a_s{\lambda^2}+{6}{\overline m_s^2}\ga
1+2.6667a_s+24.1415a_s^2+250.4705a_s^3+ {\cal O}(a_s^4)\dr\Bigg{\}}~.
\eeq
The coefficient of the $a_s^4$ term like the ones of
all unknown higher order terms will be mimiced by the $\lambda^2$-term \cite{ZAK,CNZ,SNV,CN,SN05} 
present in the $D=2$ and $D=4$ contributions.
The presence of $\lambda^2$ in the OPE helps in resolving the old puzzle of hierarchy scale \cite{NSVZ} 
encountered in the sum rules analysis of the
pion \cite{PSEUDO,PSEUDO2,SNL,SNB} and gluonia \cite{SNGLUE} channels. 
$\lambda^2$ also improves the determination of $\alpha_s$ and $m_s$
from
$\tau$ decay data
\cite{SN05,SNTAU}. In the present paper, the series converge 
slowly in the region where the analysis is performed. However, we expect that this slow convergence
will not ruin the result, though introducing a large error, as each corrections are individually smaller
than the lowest order term, while the size of the $\lambda^2$ contribution (see Eq. (\ref{eq: d4}) below)
introduced to mimic the resummation of the unknown higher order terms remains a correction of the lowest order one.
Indeed, a na\"\i ve geometric estimate of the $a_s^4$ coefficient leads to a contribution of the order of 1000$a_s^4$
\cite{SNE}, which is of the same order as the one of $\lambda^2$, and then justifying the arguments which motivate its
introduction as a model for the unknown higher order terms.
\\
\nin
$\bl$ The $D=4$ contributions read \cite{BNP,SVZ,SNB}:
\bea\label{eq: d4}
 Q^4\Pi_{ss}^{(D=4)}(Q^2)&\simeq&\Bigg{\{}
\frac{1}{12\pi}\ga 1-{11\over 18}a_s\dr{\la\alpha_s G^2\ra}\nnb\\
&+&\ga 1-a_s-{13\over 3}a_s^2\dr\la 2m_s\bar ss\ra+
\ga {4\over 3}a_s+{59\over 6}a_s^2\dr\la 2m_s\bar ss\ra\nnb\\
&+&\Bigg{[}{4\over 27}a_s+\ga -{257\over 486}+{4\over 3}\zeta(3)\dr a_s^2\Bigg{]}\sum_i\la m_i\bar \psi_i\psi_i\ra\nnb\\
&+&{1\over \pi^2}\Bigg{[} -{6\over 7}+{23\over 28}a_s+\ga {731\over 56}-{18\over 7}\zeta(3)\dr a_s\Bigg{]}\bm_s^4\nnb\\
&-&{8\over \pi^2} m_s^2a_s\lambda^2
\Bigg{\}},\nnb\\
\eea
where the last term is due to the $\lambda^2$ term \cite{CNZ}, and $\zeta(3)$=1.202...\\
\nin
$\bl$ The $D=6$ contributions read \cite{SVZ}:
\bea\label{eq: d6}
 Q^6\Pi_{ss}^{(D=6)}(Q^2)=-{1\over 4\pi^2}{896\over 81}\rho\la \bar ss\ra^2~,
\eea
where $\rho\simeq 2-3$ parametrizes the deviation from the vacuum saturation assumption of the four-quark condensate.
We shall use as input $\Lambda_3=(375\pm 25)$ MeV for three flavours and
\cite{SNG,SNV,TARRACH,SNB}:
\bea\label{eq: qcd}
(m_u+m_d)\la \bar uu
+\bar dd\ra &=&-{2}m_\pi^2 f_\pi^2 \nnb \\
 (m_s+m_u)\la\bar ss+ \bar uu\ra&\simeq &
-{2}\times 0.7m_K^2 f_K^2\nnb\\
\la \bar ss\ra/ \la \bar uu\ra&\simeq& 0.7\pm 0.2\nnb\\
a_s\lambda^2&\simeq& -(0.07\pm 0.03)~{\rm GeV}^2\nnb\\
\la\alpha_s G^2\ra&\simeq&(0.07\pm 0.01)~\mbox{GeV}^4\nnb \\
\rho\alpha_s\la\bar uu\ra^2&\simeq& (5.8\pm 0.9)\times 10^{-4}~{\rm GeV}^6~,
\eea
where: $f_\pi=93.3$ MeV, $f_K=1.2 f_\pi$. We have taken into account
a possible violation of kaon PCAC as suggested by the
QSSR analysis \cite{SNK,SNB}.\\
\section{Parametrization of the spectral function}
$\bl$ For the resonances, we parametrize the spectral function within a narrow width approximation (NWA)~\footnote{A parametrization
using a Breit-Wigner form leads within the errors to the same result.} by using the most recent data compiled in PDG \cite{PDG} for the
$\phi(1019.7)$ and
$\phi'(1680)$ with:
\bea
\Gamma_{\phi(1019.7)\rar e^+e^-}\simeq (1.27\pm 0.02)~{\rm keV}~,~~~~~~~~~
\Gamma_{\phi(1680)\rar e^+e^-}\simeq (0.43\pm 0.15)~{\rm keV}~.
\eea
$\bl$ For the non-resonant contributions in the region below $\sqrt{t}\leq 1.3$ GeV, we use the sum of the exclusive rates of the
$I=0$ channel compiled in
\cite{DOLINSKY} and use a $SU(3)$ symmetry for keeping the $\phi$-component. An analogous parametrization has been used
successfully for the accurate estimate of the hadronic contribution to the muon anomalous magnetic moment \cite{SNGM2}. \\
\nin
$\bl$ Above the $\phi(1680)$, we use a QCD parametrization of the spectral function as:
\beq
{1\over \pi}{\rm Im} \Pi_{ss}(t\geq t_>)\simeq \theta (t-t_>){1\over 4\pi^2}
\Bigg{\{}1+a_s(t)(1+2m_s^2/t)+1.6398a_s^2(t)-10.284a_s^3(t)\Bigg{\}}~,
\eeq
where $t_>$ is the QCD continuum threshold.
\section{FESR}
We shall use the combination of FESR introduced recently in \cite{SNTAU} for extracting $m_s$ from the $V+A$ component of the $\tau$ decay data
\footnote{More technical details of this sum rule can be found in \cite{SNTAU}.}:
\bea\label{eq:combine}
{\cal S}_{10}&\equiv& {\cal M}_{0}-{2\over t_c}{\cal M}_{1}\equiv \int_0^{t_c} dt \ga 1-2{t\over t_c}\dr
\frac{1}{\pi}{\rm Im}\Pi_{ss}(t)~,
\eea
which is sensitive, to leading order, to $m^2_s$ and $\lambda^2$. Unlike the individual sum rules, these
combinations of sum rules are less sensitive to the high-energy tail of the spectral functions (effect of
the $t_c$-cut), as it is chosen such that at $t=t_c$ the integral vanishes. 
\subsection*{Test of duality}
\begin{figure}[H]
\begin{center}
\includegraphics[width=7cm]{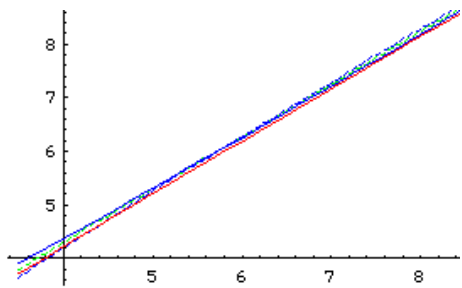}  
\caption{\footnotesize FESR prediction of $t_c$ versus $t_c$ in GeV$^2$. The green curve corresponds the central value of the data;
the blue ones to the larger and smaller values of the data; the red line is the solution $t_c=t_c$, where both beyond this
value, the predictions from the LHS and RHS should exactly co\"\i ncide. The curves correspond to the value $t_>=1.75$ GeV$^2$;
$\hat m_s=(56-145)$ MeV and
$a_s\lambda^2=-(0.04-0.1)$ GeV$^2$.}
\label{fig: dual}
\end{center}
\end{figure}

In principle, the value of the $t_c$-cut of the FESR integrals is a free parameter. We fix its optimal value by looking
for the region where the phenomenological and QCD sides of the ratio of moments:
\bea\label{eq: ratio}
{\cal R}_{10}\equiv 2{{\cal M}_{1}\over {\cal M}_{0}}~, 
\eea
are equal. We present this analysis in Fig. \ref{fig: dual}, by showing the value of $t_c$ predicted by the sum rule versus
$t_c$ and by comparing the result with the exact solution
$t_c=t_c$ expected to hold for all values of $t_c$ because the continuum is paramterized by the QCD expression above $t_c$.
From Fig. 1, one can deduce, that QCD duality is best obtained at:
\beq
t_c\simeq (6.0\pm 0.5)~{\rm GeV}^2~.
\eeq
\nin
where one expects to get the optimal value of $m_s$ from FESR. In order to get this number, we have
used the value of the invariant mass
$\hat{m}_s=(56-145)$ MeV, which is like $\lambda^2$ a tiny correction in this duality test analysis.
Once we have fixed the value of $t_c$ where the best duality from the two sides of FESR has been obtained, we can now
estimate $m_s$.
\subsection*{Estimate of   $\hat m_s$ versus  $\lambda^2$}
In principle, $t_>$, the beginning of the QCD continuum parametrization is a free parameter. We study the stability of the
$m_s$ output  against the  $t_>$-variation and found that in the range $\sqrt{t_>}=(1.75\pm 0.05)$ GeV, $m_s$ is quite stable. We
present the result of the invariant mass $\hat m_s$ for different values of $\lambda^2$ in Table \ref{tab:
comparison}. 
\begin{table*}[H]
\setlength{\tabcolsep}{1.5pc}
\begin{center}
\caption{$\hat m_s$ versus $\lambda^2$ for $t_c=6$ GeV$^2$ and $t_>=1.75$ GeV$^2$.}
\label{tab: comparison}
\begin{tabular}[hbt]{lc}
&\\
\hline
\hline
\\
$-a_s\lambda^2$ in GeV$^2$&$\hat m_s$ in MeV \\
\\
\hline
\hline
\\
0.02& $77\pm 55\pm 15\pm 8$\\
0.04& $94\pm 36\pm 10\pm 7$\\
0.06&$108\pm 30\pm 9\pm 6$\\
0.07&$114\pm 27\pm 9 \pm 6$\\
0.08&$120\pm 25\pm 6\pm 5$\\
0.10&$131\pm 22\pm 5\pm 4$\\
0.12&$142\pm 21\pm 5\pm 4$\\
&\\
\hline
\hline
\end{tabular}
\end{center}
{\footnotesize 
\begin{quote}
The 1st error is due to the data, the 2nd one to $t_c$ and $t_>$, and the 3rd one to the QCD parameters. 
\noindent
\end{quote}}
\end{table*}
\nin
Using the
value of $a_s\lambda^2$ given in Eq. (\ref{eq: qcd}), we deduce the predictions:
\beq\label{eq: fesr}
\hat m_s=(114\pm 27\pm 9 \pm 6\pm 19)~{\rm MeV}~~~\Longrightarrow~~~ {\overline m}_s(2~{\rm GeV})=(100\pm 28_{\rm exp}\pm 20_{\rm th})~{\rm MeV}~,
\eeq
where the last error in $\hat m_s$ is due to $\lambda^2$ \footnote{We have taken the central value of the asymmetrical
errors due to $\lambda^2$.}. We have used the conversion scale:
\beq
\bm_s(2~{\rm GeV})\simeq 0.876~ \hat m_s~. 
\eeq
One can notice that the result is perfectly consistent with
the one from $\tau$-decay \cite{SNTAU} given in Table~\ref {tab: compare} using the same combination of FESR. Like in the $\tau$-decay
data, the  theory with $\lambda^2=0$ tends to give too small value of $m_s$ though the error is quite large.
\section{$ \tau$-like sum rules revisited}
$\tau$-like sum rules have been proposed in SN \cite{SNE} for extracting $m_s$ from $e^+e^-$ data, which have been exploited later
on in \cite{MALT,JAMINE} using the same or a slight variant of the SN-sum rule. SN \cite{SNE} results have been confirmed by \cite{JAMINE} using the
expected small effects of the
$SU(2)$ breaking due to $\omega-\rho$ mixing. However, the central value of the results obtained in \cite{SNE,JAMINE} are slightly higher than recent
estimates, though consistent within the errors. In the following, we reconsider the original sum rules:
\beq
R_{\tau,\phi}\equiv\frac{3|V_{ud}|^2}{2\pi\alpha^2}S_{EW}\int_0^{M^2_\tau}
ds\ga 1-\frac{s}{M^2_\tau}\dr^2\ga 1+\frac{2s}{M^2_\tau}\dr\frac{s}{M^2_\tau}
\sigma_{e^+e^-\rar \phi,\phi',...}~,
\eeq
and the $SU(3)$-breaking combinations:
\beq
\Delta_{1\phi}\equiv R_{\tau,1}-R_{\tau,\phi}.
\eeq
Here, $S_{EW}=1.0194$ is the electroweak correction \cite{MARC} and $|V_{ud}|^2=0.975$ is the CKM mixing angle. The QCD expressions of these sum rules
have been given in
\cite{SNE}, except that we shall replace the contributions of the uncalculated $a_s^4$ and HO terms of the PT series by the contribution of
$\lambda^2$. We shall also use the computed coefficient of the $a_s^3m_s^2$ term $k_3=250.4$ \cite{CHETKUHN} instead of the
estimated 218.55 used in~\cite{SNE}. Contrary to the previous sum rules, the $\tau$-like sum rule is more precise near the real
axis due to the presence of the threshold factor $\ga 1-{s}/{M^2_\tau}\dr^2$. 
\subsection*{Upper bound on  $m_s$ from   $R_{\tau,\phi}$} 
\begin{figure}[H]
\begin{center}
\includegraphics[width=8cm]{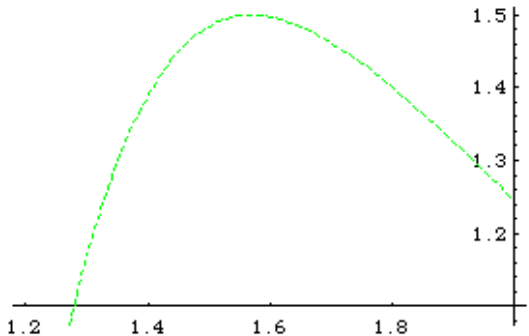}  
\caption{\footnotesize Upperbound of $R_{\tau,\phi}$ versus $M_\tau$ in GeV for
$a_s\lambda^2=-0.07$ GeV$^2$, using  the central value of the data.}
\label{fig: bound}
\end{center}
\end{figure}

We shall use the positivity of $R_{\tau,\phi}$ and
saturate the spectral function by the
$\phi$(1019.7) contribution. In this way, we derive a lower bound on $R_{\tau,\phi}$ which we show in
Fig. \ref{fig: bound} versus $M_\tau$. 
\nin
From this figure, one can derive:
\beq\label{eq: bounde+e-}
\hat m_s\leq(172\pm 16)~{\rm MeV}~~~\Longrightarrow~~~ {\overline m}_s(2~{\rm GeV})\leq (151\pm 14)~{\rm MeV}~.
\eeq
This upper bound is comparable with the previous value ${\overline m}_s(2~{\rm GeV})\leq (147\pm 21)~{\rm
MeV}$ obtained in \cite{SNE} using the same method, where the improved accuracy comes mainly from a more precise value of
the $\Gamma_{\phi\rar e^+e^-}$ width. This bound is also comparable with the one 148 MeV from a direct estimate of the quark
chiral condensate \cite{BOUND2}.
\subsection*{Extraction of $  m_s$ from $  R_{\tau,\phi}$}
Parametrizing the spectral function by the
$\phi(1019.7),~\phi(1680)$ and the non-resonant contributions below 1.39 GeV, we deduce, for the QCD parameters in Eq. (\ref{eq:
qcd}), the results in Table \ref{tab: rphi}. Comparing these results with the previous ones in \cite{SNE}, we notice that the inclusion of the
non-resonant states contributions has increased slightly the value of $R_{\tau,\phi}$. However, this small change affects the value of
$m_s$ which is also a correction in the QCD expression of $R_{\tau,\phi}$. 
\begin{table*}[hbt]
\begin{center}
\setlength{\tabcolsep}{2.2pc}
\caption{Phenomenological estimates of $R_{\tau,I}$ and central
values of $\hat{m}_s$ to order $\alpha_s^3$}
\label{tab: rphi}
\begin{tabular}{c c c}
&&\\
\hline 
\hline
 &&\\
$M_\tau$ &$\frac{9}{2}R_{\tau,\phi}$&
$\hat{m}_s$\\ &&\\
\hline
\hline
&&\\
1.4&$2.82\pm 0.47$&--\\
1.6&$1.73\pm 0.05$&111\\
1.7&$1.66\pm 0.05$&128\\
1.8&$1.65\pm 0.06$&129\\
1.9&$1.65\pm 0.06$&127\\
2.0&$1.80\pm 0.11$&--\\
&&\\
\hline 
\hline
\end{tabular}
\end{center}
\end{table*}
Taking as an optimal estimate of $m_s$ the value which is stable in the change of $M_\tau\simeq (1.8\pm 0.1)$ 
GeV~\footnote{Notice that the inclusion of higher states has slightly shifted the position of the optimum from 1.6 GeV (Fig.~\ref{fig:
bound}) to 1.8 GeV (Table~\ref{tab: rphi}).}, we obtain:
\beq\label{eq: rphi}
\hat m_s=(129\pm 15_{\rm exp}\pm 25_{\rm th})~{\rm MeV}~~~\Longrightarrow~~~ {\overline m}_s(2~{\rm GeV})=(113\pm
13_{\rm exp}\pm 22_{\rm th})~{\rm MeV}~.
\eeq
This value is consistent with the result in Eq. (\ref{eq: fesr}). Like the previous
FESR, this prediction is also affected to leading order by $\lambda^2$, where the value of $m_s$ increases with $-a_s\lambda^2$.
For $\lambda^2=0$, the corresponding value of $\hat m_s$ is 104 MeV, which is still
consistent with the other determinations, though in the lower side.  Therefore, one can notice that, contrary to the $\tau$-sum rule,
this sum rule cannot differentiate between the two cases $\lambda^2=0$ and $\lambda^2\neq 0$.\\
\subsection*{Extraction of   $m_s$ from $ \Delta_{1\phi}$}
Here, we analyze the $\Delta_{1\phi}$ sum rule. Unlike $R_{\tau,\phi}$, $\Delta_{1,\phi}$ is not sensitive to leading order to
$\lambda^2$. However, like the $SU(3)$-breaking sum rule, it has the inconvenience to involve the difference of two large
independent channels (isovector-isoscalar here). Using the optimal value of
$R_{\tau,\phi}$ at
$M_\tau=1.8$ GeV from Table \ref{tab: rphi}, and using the value of $R_{\tau,1}=1.78\pm 0.025$ at this scale from $\tau$-decay data \cite{EXP}, one can deduce:
\beq
\Delta_{1\phi}(1.8~{\rm GeV})=0.13\pm 0.06~,
\eeq
leading to:
\beq\label{eq: delphi}
\hat m_s=(113\pm 26_{\rm exp}\pm 4_{\rm th})~{\rm MeV}~~~\Longrightarrow~~~ {\overline m}_s(2~{\rm GeV})=(99\pm
23_{\rm exp}\pm 4_{\rm th})~{\rm MeV}~,
\eeq
which is in good agreement with the former determinations.
\subsection*{Final value of $m_s$ from $e^+e^-$}
One can see in Table \ref{tab: compare}, that there are good agreement between results from $e^+e^-$ data from
alternative forms of the sum rules, indicating the reliability of the results. Taking the average of the results in Eqs. (\ref{eq:
fesr}), (\ref{eq: rphi}) and (\ref{eq: delphi}), we deduce the final value from
$e^+e^-$ data:
\beq\label{eq: e+e-}
\hat m_s=(119\pm 17)~{\rm MeV}~~~\Longrightarrow~~~ {\overline m}_s(2~{\rm GeV})=(104\pm
15)~{\rm MeV}~.
\eeq
This value is consistent within the errors (though in its higher side) with the old value: 
${\overline m}_s(2~{\rm GeV})$ =$(125\pm 14_{\rm exp}\pm 20_{\rm
th})$ MeV in \cite{SNE}, and the one $(139\pm 31)$ MeV in \cite{JAMINE}. 
\section{Present status of light quark masses}
In this part, we present the status of the recent determinations of the light quark masses from different sum rule channels, to
the same order
$\alpha_s^3$ and including the $\lambda^2$ term, which we also compare with the lattice results including dynamical fermions. 
\subsection*{The (pseudo)scalar channels}
These channels are, in principle, the best place for extracting the value of the light quark masses because these masses enter as the
leading overall coefficients in the corresponding QCD correlators, precisely known up to order $\alpha_s^4$ or alternatively
to order $\alpha_s^3$ plus the $\lambda^2$-term which mimics the unknown higher order terms.\\
$\bl$ Estimates of the sum of the light quark masses $(m_u+m_d)$ \cite{PSEUDO,PSEUDO2} have been updated in \cite{CNZ,SNL,SNB} by
including the
$\lambda^2$ and $a_s^3$ corrections, and by using the parametrization of the $3\pi$ spectral function \cite{PSEUDO2} which satisfies
the ChPT constraints. The result is:
\beq\label{eq: pseudo}
(\bm_u+\bm_d)(2~{\rm GeV})\simeq (8.6\pm 2.1)~{\rm MeV}~, 
\eeq
where $\lambda^2$ decreases the value of the sum by 5$\%$. Combined with the ChPT ratio \cite{GASSER}:
\beq\label{eq: chpt}
r_3\equiv {2m_s\over (m_u+m_d)}=24.4\pm 1.5~,
\eeq 
one can deduce the value of $m_s$ given in Table \ref{tab: compare} \footnote{Kaon sum rule also gives an analogous value
\cite{PSEUDO,SNL,SNB} but the corresponding spectral function is less controlled by ChPT than the one of the pion.}.\\
$\bl$ Direct extractions of $m_s$ from kaon sum rules also exist in the literature \cite{MALT2,CHET2}. Here, the analysis
suffers from the unmeasured value of the kaon radial excitations $K(1460)$ and $K(1830)$ decay constants which play an important role at the
scale where the sum rules are optimized. We expect that the errors induced by this model dependence have not yet been properly included in
the quoted small errors of the estimated decay constants. 
\\
\nin
$\bl$ The scalar channel has been revisited in \cite{OLLER} using $K\pi$ phase shift data \footnote{For some
recent discussions on the scalar channels from the sum rules, see e.g. \cite{SNS}.}, with the resulting value of the quark
mass given in Table
\ref{tab: compare}. Here, the phenomenological side of scalar sum rule is better known due to the availability of the
$K\pi$-phase shift data combined with the constraints from ChPT. \\
\nin
$\bl$ Lower bounds on the light quark masses have been also derived from the (pseudo)scalar channels using the positivity of the spectral
function \cite{BOUND}. These bounds have been updated in \cite{SNL,SNB} by including the effect of $\lambda^2$ and the order $\alpha_s^3$ PT
contributions. The best updated bounds from pseudoscalar kaon and pion sum rules given in \cite{SNL,SNB} including the
$\alpha_s^3$ and $1/q^2$-term are:
\beq\label{eq: pseudobound}
 \bm_s(2~{\rm GeV})\geq (71\pm 4)~{\rm MeV}~,~~~~~~~~~~~~ (\bm_u+\bm_d)(2~{\rm GeV})\geq (5.9\pm 0.3)~{\rm MeV}~.
\eeq
The inclusion of the $\alpha_s^4$ term obtained in \cite{CHET3} decreases this value by about 2 MeV. However, slightly
different central values without error bars have been given in \cite{CHET3} but agree with ours within our quoted errors.
\subsection*{$  \tau$-decay data}
$\tau$-decay data have been used using different methods for extracting $m_s$: \\
\nin
$\bl$ In \cite{PICH,KUHN}, $SU(3)$-breaking moment sum rules involving
the difference of the non-strange and strange $V+A$ components of $\tau$-decay data have been used. These sum rules have the
advantage to be non-affected by $\lambda^2$ to leading order, but have the inconvenience to involve a strong cancellation of the two independent
channels $\bar ud$ and $\bar us$. The value of $m_s$ is given in Table \ref{tab: compare}.\\
\nin
$\bl$ Alternatively, a combination of FESR involving only the $\Delta S=-1$, but sensitive to $\lambda^2$ to leading order, has been
proposed in
\cite{SNTAU}. This sum rule has been used for studying the effect of $\lambda^2$ on the value of $m_s$. Giving the value of $\lambda^2$ in
Eq. (\ref{eq: qcd}), one obtains the value in Table
\ref{tab: compare}.
\begin{table*}[H]
\setlength{\tabcolsep}{1.5pc}
\begin{center}
\caption{Recent phenomenological determinations of $\overline{ m}_s$ (2 GeV) to order $\alpha_s^3$ including
the tachyonic gluon mass
$\lambda^2$ which parametrizes the UV renormalon contributions into the PT series.}
\label{tab: compare}
\begin{tabular}[hbt]{lcc}
&\\
\hline
\hline
\\
Channels&Refs.&$\bm_s$(2 GeV) in MeV \\
\\
\hline
\hline
\\
$ \bf e^+e^-$ {\bf data} \\
FESR&This work& $100\pm 28_{\rm exp}\pm 20_{\rm th}$\\
$R_{\tau,\phi}$&This work& $113\pm
13_{\rm exp}\pm 22_{\rm th}$\\
$\Delta_{1\phi}$&This work&$99\pm
23_{\rm exp}\pm 4_{\rm th}$\\
\it Average& \it This work& $  \it 104.3\pm 15.4$\\
\\
$\bf\tau${\bf -decay data}\\
FESR&\cite{SNTAU}&$93\pm 30$\\
$SU(3)$ breaking SR&\cite{PICH,KUHN}&$81\pm 22$\\
\\
{\bf Pseudoscalar }\\
Pion SR + ChPT&\cite{CNZ,SNL,SNB}&$105\pm 26$\\
Kaon FESR&\cite{MALT2}&$100\pm 12$\\
Kaon Exponential SR&\cite{CHET2} $^{+)}$&$103\pm 9$\\
\\
{\bf Scalar }\\
$K^*_0$ SR&\cite{OLLER} $^{*)}$&$88\pm 8$\\
\\
{\bf Chiral condensate}\\
$N$ , $B^*-B$, $D\rar K l\nu$ &\cite{BOUND2,SNL,SNB} $^{**)}$&$131\pm 18$\\
&\\
\hline\\
{\bf Final  Average  $ ^{\dagger)}$}& Weighted&$ 96.10\pm 4.80$\\
& Arithmetic&$ 96.30\pm 17.5$\\
\\
\hline
\hline
\end{tabular}
\end{center}
{\footnotesize 
\begin{quote}

$^{+)}$ Result at order $\alpha_s^3$ quoted in \cite{CHET2}; $\alpha_s^4$ corrections
increase the value by 2 MeV.\\
$^{*)}$ Result to order $\alpha_s^4$ quoted in \cite{OLLER}. $\alpha_s^4$ correction is expected like in the pseudoscalar channel to be
small.\\
$^{**)}$ Not included in the average as known to order $\alpha_s$.\\
$^{\dagger)}$ We have assumed that the different determinations are independent from each others.
\noindent
\end{quote}}
\end{table*}
\nin
\subsection*{Direct extraction of the chiral condensate  $ \la \bar \psi\psi\ra$}
Extraction of the chiral condensate $ \la \bar \psi\psi\ra$ has been used in \cite{BOUND2} for estimating and bounding $m_s$. The lower and upper bounds
The nucleon and $B^*-B$ sum rules give \cite{BOUND2}:
\beq
\la\bar\psi\psi\ra(M_N)\simeq[-(225\pm 9)~{\rm MeV}]^3~,
\eeq
which combined with the GMOR relation and the ChPT mass ratio in Eq. (\ref{eq: chpt}) leads to:
\beq
(\bm_u+\bm_d)(2~{\rm GeV})\simeq (10.8\pm 1.3)~{\rm MeV},~~~~~{\rm and}~~~~~\bm_s(2~{\rm GeV})\simeq (131\pm 18)~{\rm MeV}~.
\eeq
Bounds have been also derived from the $D\rar K^*l\nu$ decays \cite{BOUND2}:
\beq
0.6\leq \la\bar\psi\psi\ra(1~{\rm GeV})/[-229~{\rm MeV}]^3\leq 1.5~,
\eeq
giving:
\beq
6.8~{\rm MeV}\leq (\bm_u+\bm_d)(2~{\rm GeV})\leq 11.4~{\rm MeV}~.
\eeq
Combined with the ChPT mass ratio $r_3$ in Eq. (\ref{eq: chpt}),
 it gives:
\beq\label{eq: chiral}
82~{\rm MeV}\leq \bm_s(2~{\rm GeV})\leq 138~{\rm MeV}~,
\eeq
which are comparable with the lower bound from the pseudoscalar SR given in Eq. (\ref{eq: pseudobound}) and upper bound from $e^+e^-$ data given
in Eq. (\ref{eq: bounde+e-}). However, as the result is obtained to order $\alpha_s$, we will not consider these bounds in our final estimate. Instead, we use the allowed region in order to deduce the inaccurate estimate given in Table \ref{tab: compare}. 

\subsection*{Final value of the strange quark mass and $ m_b/m_s$ from QSSR}
$\bl$ As a final result, we consider the weighted average given in Table \ref{tab: compare} which emphasizes the
contributions of the most accurate results from (pseudo)scalar channels, which give more weight in the averaging procedure:
\beq\label{eq: msfinal}
\bm_s(2~{\rm GeV})= (96.10\pm 4.80)~{\rm MeV}~.
\eeq
$\bl$ As discussed in previous subsection, the
precisions from these two channels can be qualitatively understood because the square of the strange quark mass enters as the leading
overall coefficient in the analysis of the correlator associated to the divergence of the axial-vector [resp. of the vector] currents, while in
the vector and tau-decay channels $m_s$ enters as 
$m_s^2/q^2$ corrections in the corresponding two-point correlator.  The accuracy of the phenomenological side of the pseudoscalar sum rule is more questionable due to the lack of data
and to the accuracy of the radial excitation decay constants which play a crucial role in the analysis. 
The phenomenological side of scalar sum rule is in a better shape due to the availability of the $K\pi$-phase shift data and to the
constraints from ChPT.  A confirmation of the accuracy obtained from the (pseudo)scalar channels requires an independent analysis of
these channels.\\
$\bl$ However, it is difficult to quantify with a good precision the systematic errors of the different sum rules approach,
though, in each analysis, the different authors have used their own estimate of such errors by studying the effects of external
parameters (sum rule scale, continuum threshold,...) based on optimization or/and stability procedures or duality tests. Due to the
remarkable  good agreement of the different results given in Table \ref{tab: compare} within about 1 $\sigma$, we might expect that the quoted error in Eq. (\ref{eq:
msfinal}) is quite realistic though relatively small. This value of $m_s$, is consistent with the older  sum rule 
(arithmetic) average $(117.4\pm 23.4)$ MeV quoted in
\cite{SNB,SNE}, though in the lower side, indicating the stability of the sum rule results with time and then their reliability.\\ 
$\bf$ The more conservative value from the arithmetic average obtained in Table \ref{tab: compare}:
\beq\label{eq: arithmetic}
\bm_s(2~{\rm GeV})=~{\rm (96.3\pm 17.5)~ MeV},
\eeq
can be translated  into  the range of  $m_s$ values allowed by the sum rules analysis :
\beq\label{eq: range}
79~{\rm MeV}\leq \bm_s(2~{\rm GeV})\leq{\rm 114~ MeV},
\eeq
which can be compared with the rigorous lower [resp. upper] bound coming from the positivity of the spectral functions in the  pseudoscalar [resp. $\phi$] sum rules updated to order $\alpha_s^3$~\footnote{Stronger lower and
upper bounds from a direct extraction of the chiral condensate has been obtained in Eq.~(\ref{eq: chiral}), but they are only known to order $\alpha_s$.}:
\beq\label{eq: boundfinal}
(71\pm 4)~{\rm MeV}\leq \bm_s(2~{\rm GeV})\leq{\rm (151\pm 14)~ MeV}.
\eeq
$\bl$ 
The final average value of $m_s$ obtained from phenomenological methods given in Eq. (\ref{eq: msfinal})  are inside the range of values quoted by PDG
\cite{PDG}:
\beq\label{eq: latt1}
\bm_s(2~{\rm GeV})= (80\sim 130)~{\rm MeV}~.
\eeq
$\bl$ One can also compare the value in Eq. (\ref{eq: msfinal})  with the different
lattice results \cite{LATT,LATT2,LATT3}.  The recent lattice results including dynamical fermions are:
\beq\label{eq: latt2}
\bm_s(2~{\rm GeV})\vert_{n_f=2}= (100\sim 130)~{\rm MeV}~,~~~~~~~~~~~~~~~~\bm_s(2~{\rm GeV})\vert_{n_f=2+1}= (70\sim 90)~{\rm MeV}~,
\eeq
which appear to depend on the number of flavours. The difference of the results for $n_f=2$ \cite{LATT2} and $n_f=2+1$
\cite{LATT3} (see however \cite{ISHIKAWA}) and the slightly higher prediction of the ChPT mass ratio
$r_3=27.4\pm 4.2$ defined in Eq. (\ref{eq: chpt}) may indicate that, it is premature, at present, to extract a precise value of $m_s$ from
the lattice calculations, before a reliable control of the systematic errors, higher order terms and some other effects.\\
$\bl$ Running $\bm_s$ until $\bm_b(\bm_b)=(4.23\pm 0.06)$ GeV \cite{PDG,SNB,SNL,SNMB}, by taking care on the threshold effects, one can
deduce the useful scale-independent quantity for model-buildings:
\beq\label{eq: ratio5}
r_5\equiv {m_b\over m_s}= 50 \pm 3~.
\eeq
\subsection*{Implied values of the up and down quark masses from QSSR +ChPT}

$\bl$ Using the previous value of $\bm_s(2~{\rm GeV})$ in Eq. (\ref{eq: msfinal}) together with the ChPT mass
ratio in Eq.~(\ref{eq: chpt}), we can deduce:
\beq\label{eq: mud}
(\bm_u+\bm_d)(2~ {\rm GeV})=(7.9\pm 0.6)~{\rm MeV}~,
\eeq
in nice agreement with the result from the pion sum rule in Eq. (\ref{eq: pseudo}).
Using again the ChPT mass ratio \cite{GASSER} \footnote{The different values of $ (m_u+m_d$) and  $
(m_d-m_u$) from respectively the pseudoscalar and scalar sum rules
\cite{PSEUDO,SCAL,SNL,SNB} excludes the possibility to have $m_u=0$.}:
\beq\label{eq: chpt2}
{m_u\over m_d}= 0.553\pm 0.043~,
\eeq
we obtain:
\beq \label{eq: mu}
\bm_d(2~ {\rm GeV})=(5.1\pm 0.4)~{\rm MeV}~,~~~~~~~~~~~~~~~~ \bm_u(2~ {\rm GeV})=(2.8\pm
0.2)~{\rm MeV}~.
\eeq
$\bl$ Taking the average value of $\bm_s$ (2 GeV)= $(95.8\pm 4.9)$ MeV, by excluding the pion sum rule result in Table
\ref{tab: compare}, and using the prediction of
$(m_u+m_d)$ from the pion sum rule in Eq. (\ref{eq: pseudo}), one can deduce the ratio:
\beq\label{eq: ratio3}
r_3\equiv {2m_s \over (m_u+m_d)}=23.5\pm 5.8~,
\eeq
in perfect agreement with the ChPT mass ratio in Eq. (\ref{eq: chpt}). 

\section{Conclusions}
We have revisited the estimate of the strange quark mass from $e^+e^-$ data. Including the $\alpha_s^3$ plus a
phenomenological estimate of the UV renormalon contributions parametrized by the tachyonic gluon mass $\lambda^2$, we deduce the final value
from $e^+e^-$ data in Eq. (\ref{eq: e+e-}) and  reported in Table~\ref{tab:
compare}. We compare this value with recent determinations from different channels in Table~\ref{tab: compare} known to the same level of
approximations.  Our final result coming from a weighted average of different determinations from
Table~\ref{tab: compare} is given in Eq.~(\ref{eq: msfinal}).  The updated lower and upper bounds for the strange quark mass to order
$\alpha_s^3$ are summarized in Eq. (\ref{eq: boundfinal}). 
The value and range of $m_s$ given in Eq. (\ref{eq: msfinal}) to Eq. (\ref{eq: boundfinal}) are
inside the PDG values quoted in Eq. (\ref{eq: latt1}) and agree within the errors with the recent lattice calculations in Eq.~(\ref{eq: latt2})
including dynamical fermions. 
Combining  the final average result of $m_s$ in Eq. (\ref{eq: msfinal}) with the ChPT mass ratios, we deduce the
value of the running $u$ and
$d$ quark masses in Eqs.~(\ref{eq: mud}) and (\ref{eq: mu}), while we also predict in Eqs. (\ref{eq: ratio3}) and (\ref{eq: ratio5}), the
useful scale-independent mass ratios $m_s/(m_u+m_d)$ and $m_b/m_s$. 

\end{document}